\documentstyle[aps,times,amsmath]{revtex}
\draft
\input epsf
\epsfverbosetrue

\begin{document}

\title{The Electrochemical Carbon Nanotube Field-Effect Transistor}
\author{ M.\ Kr\"uger, M.\ R.\ Buitelaar, T.\ Nussbaumer, and
 C.\ Sch\"onenberger$^1$}
\address{
  Institut f\"ur Physik, Universit\"at Basel,
  Klingelbergstr.~82, CH-4056 Basel, Switzerland}
\author{ L.\ Forr\'o}
\address{Institut de G\'enie Atomique, \'Ecole Polytechnique
F\'ed\'erale de Lausanne, CH-1015 Lausanne, Switzerland}
\date{\today}
\maketitle

\begin{abstract}
We explore the electric-field effect of carbon nanotubes (NTs)
in electrolytes. 
Due to the large gate capacitance,
Fermi energy ($E_F$) shifts of order \mbox{$\pm 1$\,V} 
can be induced, enabling to tune NTs
from p to n-type. Consequently, large 
resistance changes are
measured. At zero gate voltage the NTs
are hole doped in air with
\mbox{$|E_F|\approx 0.3\dots 0.5$\,eV}, corresponding to
a doping level of \mbox{$\approx 10^{13}$\,cm$^{-2}$}.
Hole-doping increases in the electrolyte. 
This hole doping (oxidation) 
is most likely caused by the 
adsorption of oxygen in air \cite{Gas-FET} and 
cations in the electrolyte.
\end{abstract}

\vspace{.3cm}



Carbon nanotubes (NTs), in particular single-wall NTs (SWNTs),
are prototype one-dimensional ($1$d) conductors, which 
ideally come two
forms, either as metals or semiconductors.\cite{BASICS} 
This classification assumes, that NTs are undoped.
An important parameter is the 
position of the Fermi energy $E_F$ 
(the chemical potential) with respect to the 
charge neutrality point (CNP). 
For an {\em undoped} NT $E_F$
coincides with the CNP \mbox{($E_F=0$)}.
Electron (n) or hole (p) doping shifts
the Fermi energy up or downwards. If the doping 
induced Fermi level
shifts are larger than the energy separation
between the $1$d-subbands, a semiconducting NT is
turned into a metallic one. 
In previous work on SWNTs
the characteristic 
$1$d density-of-states (DOS) 
was measured,\cite{STM-SPECTROSCOPY} from which $E_F<0$
was deduced.\cite{Venema2000Datta1999} 
Hole-doping was also inferred from NT-based
field-effect transitors.\cite{NT-FET,NT-FET-Avouris,Dai2000}
In contrast to semiconducting SWNTs, 
only weak field effects were 
observed in MWNTs.\cite{NT-FET-Avouris}
There are also some early measurements on thin films,
which suggest that MWNTs are hole 
doped, too.\cite{HallEffect}

This letter reports on a new gating method, electrochemical
gating, which is so effective that $E_F$ can 
be determined unambiguously
on a single MWNT. An extreme sensitivity
of the net doping concentration on the environment, in our 
case different electrolytes, is observed. Because the
doping is reflected in the measured electrical
resistance, nanoscaled sensors, such as pH sensors can
be envisaged.

Electrochemical gating is studied on single MWNTs
with lithographically defined Au contacts
evaporated over the NTs
(Fig.~1a).\cite{ApplPhysA}
The nanotube-contact structure is fabricated on
degenerately doped Si with a \mbox{$400$\,nm}
thick \mbox{SiO$_2$} spacer layer.
The Si substrate can be used as a gate (`back-gate'), see
Fig.~1b.
Large changes have been observed in the
electrical resistance $R$ of SWNT-based `tube-FET's'
by using such a back-gate (BG).
The transconductance can be increased if the gate
is placed as close as possible to the NT,
ultimately into intimate contact.
This is achieved in the present work by immersing the 
NT into an electrolyte (Fig.~1c). 
The resistance of the NT-devices is 
measured on a probe-stage at room temperature.
The stage is complemented with 
a micropipette ending in a drawn glass
capillary. 
The pipette is positioned over the device and a
small droplet of size \mbox{$\alt 100$\,$\mu$m} 
is delivered. The droplet size
is chosen such that the macroscopically large bonding
pads are not immersed in the liquid
resulting in negligible
leak currents in the
resistance measurements. 
The gate contact is formed by a Pt wire
within the glass pipette. If, as schetched in Fig.~1c,
a positive gate voltage $U_g$ is applied, 
the NT-electrolyte interface is polarized
by the attraction of cations.
The gate capacitance $C_g$ is formed by the
double-layer capacitance which can be very large.
Here, we focus on experiments in
\mbox{LiClO$_4$} electrolytes, 
used at concentrations of
\mbox{$1-500$\,mM}. 

Fig.~2 compares the gate effect of a
MWNT for two cases: with
(a) liquid ion-gate and (b) BG.
While the initial electrical resistances $R$ at
$U_g=0$ are comparable,
the gate induced changes
are very different.
$dR/dU_g$  
is \mbox{$2.5$\,$\Omega$/V} in (a)
and \mbox{$570$\,$\Omega$/V} in (b). Hence, liquid-ion
gating is by a factor
\mbox{$> 200$} more effective than back gating.
Starting from \mbox{$U_g=0$}, 
$R$ increases with increasing $U_g$,
which is characteristic for p-type behavior.
With BG this increase
persists up to the largest possible gate voltages
of \mbox{$\approx 80$\,V}, where the sample is destroyed.
In contrast, $R(U_g)$ has a
maximum at \mbox{$U_g=U_0$} in the electrolyte. 
The decrease of $R$ for \mbox{$U_g \geq U_0 $}
now suggests n-type behavior.
The position of the resistance maximum therefore
marks the charge-neutrality point of the NT, i.e.
$E_F=0$, if \mbox{$U_g=U_0\approx 1$\,V}.
$R(U_g)$ is measured cyclicly. After
some cycles an equilibrium situation is established
with a relatively well defined peak position and
only weak hysteresis, provided one ramps slowly
(\mbox{$10$\,minutes} per sweep). In this example,
$R$ changes by only
\mbox{$20$\,\%}.

Fig.~3 shows another example. It illustrates the
time dependence and, most notably, shows a much larger
$R$ change.
$R_{max}$ is a factor of $5$ larger
than \mbox{$R(0)$}. Of all our measured samples, 
approximately half display a weak $R$ change of
order \mbox{$20-50$\,\%}, whereas $R$ changes
by several \mbox{$100$\,\%} for the other half. 
The first up-sweep (increasing $U_g$)
($\times$) was followed
by a down-sweep ($+$). This is repeated until a stationary
curve is obtained ($\bullet$). It is seen that the resistance
maximum shifts to higher voltages with time to finally
reach \mbox{$U_0=1$\,V} in this case. 

In the following we will present a model which captures the
essential physics of this experiment. We assume
that only the outermost NT shell
needs to be considered\cite{AdrianAB}
and describe the NT DOS by that of
a single layer of graphite, neglecting 
$1$d~bandstructure effects.\cite{REMARK1}
Using the Einstein relation, which relates the
diffusion coefficient $D$ to the conductivity, 
the electrical conductance $G$  can be written 
as \mbox{$G=(2\pi r/L)e^2 D N_{\Box}$}. 
Here, $r$ is the radius
of the NT, $L$ the contact separation, and $N_{\Box}$ the
$2$d~DOS which depends on $E_F$.
For an ideal single sheet of graphite, the DOS
is \mbox{$N_{\Box}=2|E_F|/\pi(\hbar v_F)^2$}, where
$v_F$ is the Fermi velocity.\cite{ApplPhysA}
At the charge-neutrality point (CNP),
i.e. at \mbox{$E_F=0$}, $N_{\Box}$ vanishes.
We add a phenomenolgical parameter $E_c$
accounting for a finite DOS at the CNP due to
temperature and adsorbate induced
band-structure modifications and write
\mbox{$N_{\Box}=(2E_c/\pi(\hbar v_F)^2)(1 + (E_F/E_c)^2)^{1/2}$}.
The normalized conductance 
\mbox{$g(E_F)=G(E_F)/G(0)$}\mbox{$=(1 + (E_F/E_c)^2)^{1/2}$}
is used to fit our data. 
For this $g(U_g)$ is required,
so that the relation between $U_g$ and
\mbox{$E_F$} needs to be derived.

Fig.~4 shows schematically what happens when a NT is biased
via an external gate (engineering sign convention is used here).
There are two effects: First, there is an external
electric field $\vec{E}$ and correspondingly an
electrostatic potential difference $\phi$ between the NT and
the gate electrode. Secondly, $E_F$ 
must increase because of the addition of charge carriers to
the NT. The relations between charges 
$Q_1$, $Q_2$ (see Fig.~4c) and potentials
$E_F/e$, $\phi$ are 
determined by the geometrical
capacitance $C_g=dQ_2/d\phi$ and
chemical capacitance $C_{NT}=dQ_1/d(E_F/e)$ of 
the NT (Fig.~4c).
These two capacitors are in series.
Fig.~4b shows the energy-dependent DOS for
a general biasing condition. The externally
applied voltage $U_g$ corresponds to the electrochemical
potential $\eta$, given by 
\mbox{$eU_g=\eta=E_F + e\phi$}.
From this relation together with $dQ_1=dQ_2$
and $C_g$, $C_{NT}$ we obtain the equation
\begin{equation}
  e\frac{\partial U_g}{\partial E_F} = 1 + \frac{C_{NT}(E_F)}{C_g}~\mbox{,}
\label{equation1}  
\end{equation}
which provides us with the required relation between $U_g$ and
$E_F$. A significant simplification follows
for NTs immersed in electrolytes because
$C_{g}\gg C_{NT}$. This is shown now.
The differential NT capacitance per unit length
(denoted by $C^{\prime}$ instead of $C$)
is given by $C^{\prime}_{NT}=e^2N^{\prime}(E_F)$
with $N^{\prime}(E_F)=2\pi r N_{\Box}(E_F)$. Thus,
$C^{\prime}_{NT}=C^{\prime}_0 (1 + (E_F/E_c)^2)^{1/2}$
with $C^{\prime}_0 = 4 e^2 r E_c/(\hbar v_F)^2$.
Taking \mbox{$r=5$\,nm}, \mbox{$v_F=10^6$\,m/s}, and
\mbox{$E_c = 0.1$\,eV} one obtains
\mbox{$C_0^{\prime}\approx 100$\,pF/m}.
The gate capacitance in solution 
(the double layer capacitance)
is $C_g^{\prime}=2\pi r \epsilon/\lambda$, 
were $\epsilon$ is the dielectric constant
($\epsilon_{H_2O}\approx 80\cdot\epsilon_0$), 
$r$ the NT radius,
and $\lambda$ the screening length 
$\propto c^{-1/2}$ ($c=$\ ion concentration). 
Taking \mbox{$c=0.1$\,M},
typical numbers
are \mbox{$\lambda\approx 1$\,nm} and
\mbox{$C_g^{\prime}\approx 10$\,nF/m}.
If the NT is gated by the Si substrate a 
coupling capacitance of
\mbox{$C_g^{\prime}\approx 5$\,pF/m} 
is deduced from our experiments
(valid for \mbox{$300$\,nm} contact separation). 
Hence:
\begin{equation}
  C_g({\rm back gate}) \ll C_{NT} \ll C_g({\rm electrolyte})
\label{equation2}  
\end{equation}
If the NT is immersed in an electrolyte, the case of interest here,
the gate capacitance is much larger than
the internal NT capacitance,
and we obtain from Eqs.~\ref{equation1}
a very simple relation \mbox{$eU_g\simeq E_F$},
valid for an undoped NT. 
If it is doped,
$Q_2 \not = Q_1$. We denote the doping charge by 
\mbox{$Q_d=Q_2-Q_1$} and the 
external gate voltage required 
to induce charge neutrality
by $U_0$. Since $E_F=0$ at the CNP,
$U_0 = Q_d/C_g$. The effect of doping is
simply to shift the functional dependence of $E_F$ vs.
$U_g$, so that 
$E_F \simeq e(U_g-U_0)$.
The interpretation of the measured
two gate-sweeps is now straightforward,
because there is a one-to-one correspondence between
$U_g$ and $E_F$. $U_0$
coincides with the CNP and directly reflects
$E_F$ for an unbiased NT (in the engineering
convention $E_F>0$ corresponds to
an excess of positive carriers).
A substantial hole
doping for MWNTs immersed in \mbox{LiClO$_4$}
is evident, leading
to Fermi level shifts of 
\mbox{$\approx 1$\,eV}. What is
the origin of this considerable hole doping?

Fig.~2a shows two measurements of the same MWNT for
\mbox{$c=0.1$} and \mbox{$0.5$\,M}.
If we assume that doping is intrinsic to the NT, for example
due to defects or inclusions, the doping charge
$Q_d$ should be constant. The relation $U_0=Q_d/C_g$ predicts
that the position of the resistance maxima should shift
to lower values with increasing $c$
according to \mbox{$U_0\propto c^{-1/2}$}. Though a peak shift
in the right direction is seen in Fig.~2a, the magnitude is
far too low, suggesting that $Q_d$ 
is affected by the electrolyte itself.
This conclusion is supported by the time dependence
shown in Fig.~3.
If the NT is immersed into the
electrolyte the resistance $R$ ($U_g=0$) drops which
corresponds to a shift of $U_0$ to the right. During the first
sweep in Fig.~3, \mbox{$U_0\simeq 0.5$\,V}, whereas
\mbox{$U_0\approx 1$\,V} in all later sweeps. It is clear
that \mbox{$E_F < 0.5$\,eV} in air before immersion.
Hence, we conclude that the electrolyte induces hole doping
in the NT, the magnitude of which depends on $c$. 
Intercalation of Li-ions can be excluded
because this would lead to n-doped NTs. This leaves the
perchlorate ion \mbox{ClO$_4^-$} as the source
of doping. This (weakly) oxidizing species
seem to adsorb on the NT specifically leading to
a charge transfer which partially oxidizes the NT (hole doping).
It is evident that this oxidation is weak in the sense that
the carbon network of the NT remains intact. If the 
NT would be eroded, irreversible measurements with a
final loss of the conductance would be expected.
If \mbox{ClO$_4^-$} is able to dope NTs by physisorption
the same is expected from \mbox{O$_2$} in air. 
A large sensitivity of the NT conductance on different
kind of gases, in particular also \mbox{O$_2$}, 
have been reported very recently.\cite{Gas-FET}
This scenario is further supported
by our measurements in other electrolytes.
If a stronger oxidizing electrolyte is used, we observe an additional
shift of the $R(U_g)$ curve to the right (additional hole doping). 
In contrast, the
curve shifts to the left in a reducing solvent.

Finally, a quantitative comparison of the experiments 
with theory is possible. This is demonstrated
in the inset of Fig.~3 where a fit (taking the
full theory) to the measured conductance $G$ is shown. 
The fit yields: \mbox{$E_c\approx 0.12$\,eV} and
\mbox{$A:=C_g/C_0\approx 10$}. The product
$A E_c$ only depends on known parameters,
like $\epsilon$, $v_F$, and $c$, but not on the
NT radius $r$. Our model predicts for this product
\mbox{$0.9$\,eV} which is in good agreement 
with \mbox{$1.2$\,eV} obtained from the fit.
This agreement proves that the model of a single
tube is correct, implying that the electrical current 
flows preferentially in the outermost
shell where most of
electrical field is screened. 
The parameter $E_c$
was introduced to account for a finite DOS at the CNP.
$N^{\prime}(0)$ is found to be \mbox{$\approx 6$\,$\times$}
larger than $N^{\prime}_{1d}$ of
an ideal metallic SWNT, possibly because 
of dopant induced states.\cite{Cohen2000}
The other class of $R(U_g)$ curves,
which show a much weaker resistance change (e.g.
Fig.~2b) can be fitted too. 
However, the deduced parameters are inconsistent
with the model of a single tube. 
In these cases, the current is most likely flowing in inner
shells too, explaining the much weaker gate effect. 

For the interpretation of
previous electrical measurements,
the net doping concentration $Q_d$ 
and the Fermi-level shift for 
a`virgin' MWNT in air
are important.
The later can immediately be obtained by 
comparing $R_0$
measured in air for \mbox{$U_g=0$} with 
the $R(E_F)$ dependence of the same NT.
This is indicated
in Fig.~3: the dash-dotted line corresponds
to $R_0$ and \mbox{$\Delta U_0$} denotes
$E_F$ before immersion. Typical values
are \mbox{$0.3-0.5$\,eV}.
Comparing this with the average $1$d subband spacing
\mbox{$\hbar v_F/2d$} (\mbox{$\approx 33$\,meV} 
for a \mbox{$10$\,nm} diameter NT),
we conclude that
$9-15$ subbands may contribute to $G$ instead of 
$2$ for an ideal metallic NT.
This finding explains why previous low temperature
measurements could be fairly well described by
$2$d diffusive transport.\cite{ApplPhysA} 
A doping-induced \mbox{$E_F=0.3$\,eV}
corresponds to a doping concentration of
\mbox{$Q_d^{\prime}/e\approx 2\cdot 10^3$\,$\mu$m$^{-1}$}, or 
expressed per surface area to
\mbox{$Q_d/e\approx 0.7\cdot 10^{13}$\,cm$^{-2}$}
giving approximately one elementary charge
per $500$ carbon atoms. Finally, 
estimates for the diffusion constant $D$ can be given, too. 
We obtain \mbox{$D=170 \pm 50$\,cm$^2$/s}
corresponding to a mean-free path of \mbox{$35$\,nm},
in agreement with our previous results obtained
differently.\cite{ApplPhysA} 

MWNTs in air are hole doped
with a (sheet) doping concentration 
of \mbox{$\approx 10^{13}$\,cm$^{-2}$} caused
by the adsorption of oxygen.
If immersed in a LiClO$_4$
electrolyte doping increases further most likely
due to a specific adsoprtion of the oxidizing species
\mbox{ClO$_4^-$}.
Polarizing the NT via an electrolyte allows to move
$E_F$ over a wide range, resulting in
large resistance changes.
NTs are possibly the most sensitive FETs
for environmental application, because 
the mobile NT carriers are in intimate contact with the
environment.

This work was supported by the Swiss National Science
Foundation. We are very grateful to
J.\ Gobrecht for providing the oxidized Si substrates.



\begin{figure}[h]
  \caption{
  (a) A typical device consisting of a single MWNT
  with Au electrodes spaced by \mbox{$L=0.3\dots 2$\,$\mu$m}.
  The electrical field effect is
  studied using (b) conventional back-gating or (c)
  liquid-ion gating.  
}
\end{figure} 

\begin{figure}[h]
  \caption{
  Electrical resistance $R$ 
  of a MWNT as a function of gate voltage $U_g$
  measured (a) in a \mbox{LiClO$_4$} electrolyte for two
  ion concentrations ($0.1$ and \mbox{$0.5$\,M)} 
  and (b) in air with back gate.
}
\end{figure} 

\begin{figure}[h]
  \caption{
  Electrical resistance $R(U_g)$ 
  of a MWNT
  measured in a \mbox{$10$\,mM} \mbox{LiClO$_4$}
  electrolyte. After immersion, the measurement
  commenced at point A with the data-point sequence $\times$,
  $+$, and $\bullet$ (\mbox{$10$\,min} per curve). 
  Drawn curves are guides to the eye.
  Inset: Comparison of $G=1/R$ ($\bullet$)
  with theory (full curve).
}
\end{figure} 

\begin{figure}[h]
  \caption{
  (a) A single sheet
  nanotube is assumed to model $R(U_g)$. 
  (b) The energy-dependent DOS under a
  general biasing condition. 
  CNP denotes the charge-neutrality point,
  $\phi$, $E_F$ and $\eta$ the electrostatic, chemical,
  and electrochemical potentials. 
  (c) A geometrical ($C_g$) and chemical ($C_{NT}$)
  capacitance need to be
  considered to account for the dependence of $E_F(U_g)$.  
}
\end{figure} 


\begin{thebibliography}{99}
\bibitem{REMARK1}
This is justified, because \mbox{$k_BT$} is of order of
the subband separation at room temperature and because of
the relatively short electron-mean free path $l\alt 2\pi r$
measured in MWNTs.
\end{thebibliography}
\end{document}